\documentclass[letterpaper,twoside,12pt]{article}
\usepackage{amssymb}
\usepackage{amsmath}
\usepackage{latexsym}
\usepackage{verbatim}
\usepackage{graphicx}
\usepackage{epsfig}
\usepackage{stmaryrd}
\usepackage{rotating,graphicx}
\usepackage[english]{babel}
\usepackage{setspace}
\usepackage[english]{babel}
\usepackage[utf8]{inputenc}
\usepackage[T1]{fontenc}
\usepackage{lmodern}
\usepackage{pstricks}
\usepackage{fullpage}
\usepackage{hyperref}
\newcommand{\bra}[1]{\langle #1|}
\newcommand{\ket}[1]{|#1\rangle}

\begin{document}
\begin{large}
\begin{center}
\textbf{{\Large On the Status of Conservation Laws in Physics: Implications for Semiclassical Gravity}}
\end{center}
\end{large}

\begin{center}
\begin{large}
Tim Maudlin\\
\end{large}
\textit{Department of Philosophy, New York University, New York, USA.}\\[.5cm]
\begin{large}
Elias Okon\\
\end{large}
\textit{Instituto de Investigaciones Filos\'oficas, Universidad Nacional Aut\'onoma de M\'exico, Mexico City, Mexico.}\\[.5cm]
\begin{large}
Daniel Sudarsky\\
\end{large}
\textit{Instituto de Ciencias Nucleares, Universidad Nacional Aut\'onoma de M\'exico, Mexico City, Mexico.} \\
\end{center}


\noindent We start by surveying the history of the idea of a fundamental conservation law and briefly examine the role conservation laws play in different classical contexts. In such contexts we find conservation laws to be useful, but often not essential. Next we consider the quantum setting, where the conceptual problems of the standard formalism obstruct a rigorous analysis of the issue. We then analyze the fate of energy conservation within the various viable paths to address such conceptual problems; in all cases we find no satisfactory way to define a (useful) notion of energy that is generically conserved. Finally, we focus on the implications of this for the semiclassical gravity program and conclude that Einstein's equations cannot be said to always hold.

\onehalfspacing
\section{Introduction}
It is often the case that a principle considered essential at some stage in the development of physics continues to be perceived as such well after the \emph{philosophical} reasons for giving it a sacred character are removed from our best understanding of the physical world. This seems to be the case in regard to the status of certain conservation laws, and in particular about conservation of energy. 

In this manuscript we explore the status of energy conservation in the context of our most successful theoretical frameworks. We examine the issue within the various viable paths to address the conceptual issues within quantum theory and focus on the implications of such an analysis for the semiclassical gravity program.

In more detail, in section 2 we survey the history of the idea of a fundamental conservation law and briefly examine the role energy conservation plays in different classical contexts. In section 3 we explore the issue in quantum settings, first within the standard version of the theory and then in the context of the best-known approaches for addressing its conceptual shortcomings; in each case we evaluate the consequences of the analysis for the semiclassical approach. We end in section 4 with a brief assessment of the available paths forward.

\section{Conservation laws in classical physics}

We start by surveying the history of the idea of the law of conservation of substance. The purpose of doing so is not to provide a historical account of the development of the notion of conservation laws in physics but to expose the way in which certain conservation laws, such as energy conservation, could be wrongly perceived as metaphysical imperatives\footnote{For the history of the concept of energy and its conservation see for instance \cite{Hiebert,Lindsay,Elkana,Harman,Darrigol}.}. Next we review the modern perspective on conservation laws provided by Noether's theorem and briefly explore the role that conservation laws play in special and general relativity. 

\subsection{Why is energy conserved?}
That sounds like a reasonable question to begin with, and we will soon return to it, but that question already contains too many unargued presuppositions to be useful. Most obviously, the question presupposes that energy \emph{is} conserved. That is a claim that can reasonably be challenged, as we will see. So at the least, we should start with:

Is energy conserved?

This question is more basic than the first, but now may strike the reader as trivial. After all, the Principle of the Conservation of Energy has been a foundational principle of physics for almost two centuries. Every physicist knows how powerful the use of the Principle can be. Throw a 1 kilogram rock straight up at 1 m/s on earth, and ask how high it will go. One can, of course, plug the gravitational force into Newton's law and solve for the trajectory, then take a derivative to determine the maximum height, but how much simpler and easier to just calculate the initial kinetic energy, set it equal to the gravitational potential energy at the apogee, and solve. Throwing the rock at an angle? Just subtract the kinetic energy of the lateral motion, rinse and repeat. No one would ever want to give up such elegant shortcuts. And the elegant shortcuts \emph{work}. How could they be so practically effective if the conservation of energy were not a fundamental principle of physics?

There is an immediate rejoinder to this argument: if energy were \emph{nearly} conserved, conserved in everyday life to within some very small quantity, the practical utility of these shortcuts would be just as great. But as logically compelling as this observation is, it does not address a deeper attachment to energy conservation as \emph{the} fundamental physical principle of the universe. In a certain state of mind, even the tiniest, most practically insignificant violation of the conservation of energy is almost unthinkable. Attachment to the principle seems to run much deeper than mere empirical grounds can account for, as if the violation of the conservation law would be \emph{wrong}.

This sense of the deep foundational status of the conservation of energy poses yet another question, viz.

Why \emph{should} energy be conserved?

What is the source of this veneration of the principle, which makes any violation of it, however small, seem so momentous?

This question leads us to the last in our series, the simplest and deepest question, which sheds light on all the rest, namely

What is energy?

\subsection{The pre-Socratic view}

Western philosophy traces back to the pre-Socratic Nature-Philosophers. The pre-Socratics were a mixed bag in terms of orientation, style, and method, but pride of place for the very first philosopher in the Western tradition is commonly awarded to Thales. And the bumper-sticker version of Thales is that he held the view that All is Water. Why should this somewhat unexpected pronouncement constitute the dawn of Western philosophy and perhaps of Western theoretical physics to boot?

The key insight accorded to Thales was the notion that beneath all the complex and variegated appearances of the world lay a single common fundamental substance. Such a view requires one to largely discount the way that the world presents itself to the senses in favor of a rational theory of what is really going on. The world is not, as it appears to be, populated by completely diverse sorts of things. Rather, it is really just one sort of thing, a fundamental unity that presents itself in different superficial guises.

Why water? This is speculative, but the ancient tetrad of four elements---earth, air, fire, and water---make much more sense as a proposal for an account of the physical world if we understand them not as what we call ``elements'' but what we call ``states of matter''. On this view, earth, air, fire and water really are better understood as what we would call solid, gas, plasma, and liquid. If someone today were to say that the universe is made of solids, liquids, gases and plasmas they would not sound too far off the mark, as far as it goes. So why would Thales choose water as the single fundamental substance of the world? Because water has familiar solid (ice), liquid (water) and gaseous (steam) forms. Leaving plasmas like fire aside, Thales made a fine choice.

Thales’ lead was followed by others. Some, such as Empedocles, postulated the four classic elements separately. Others, such as Anaximander remained monists but chose a different fundamental ``stuff'' to make everything out of. Anaximander's choice was empirically unfamiliar and conceptually abstract: the \emph{apeiron} or ``indefinite'', a sort of metaphysical Silly Putty that could become everything and was nothing in particular in itself.

But all the ancients agreed on one thing that about the fundamental substance or substances: it (or they) could neither be generated nor destroyed. Democritian atoms were eternal: being uncuttable by definition they could not be destroyed. And by the rationally compelling principle that \emph{Ex Nihilo Nihil Fit} (Nothing Comes From Nothing) they could not be created either since the only other thing existing from which they could be created was the Void. It was considered to be an inviolable requirement that a fundamental substance neither be created nor destroyed, since such generation or destruction would have to be out of or into nothing. In contrast, there is no problem with non-fundamental, derivative items coming into or out of existence. To take an example from David Albert, fists can come and go at will, and the number of fists in the world satisfies no conservation law. That's because when a fist forms, nothing really fundamentally new comes into existence: all that happens is that fingers that were already in existence become rearranged.

The point is that for a certain sort of theoretical entity---a fundamental substance or substances from which all things are made---the violation of a conservation law, no matter how tiny, would be conceptually bewildering. If there is fundamentally only a certain substance or substances in the world that simply adopts different arrangements or states or forms, then the failure to be conserved appears miraculous, no matter the amount involved.

Exactly this sort of fundamental ontological role is often ascribed to energy. Before the Theory of Relativity, the conservation of energy was accepted as a fundamental physical principle: the First Law of Thermodynamics. But after the equation $E = mc^2$ the situation got much more far-reaching. It is tempting to read that iconic equation as saying that matter (or more properly mass) \emph{just is} energy.
For example, Einstein stated\footnote{Transcription of an Einstein lecture downloaded from \url{http://www.pitt.edu/~jdnorton/teaching/HPS_0410/chapters/E=mcsquared/voice1.mp3}.}: ``It followed from the Special Theory of Relativity that mass and energy are both manifestations of the same thing, a somewhat unfamiliar conception for the average man. Furthermore, the equation $E$ is equal $mc^2$, in which energy is set equal to mass multiplied with the square of the velocity of light, showed that very small amount of mass may be converted into a very large amount of energy.'' An atomic nucleus is a certain amount of energy tied into a knot, as it were. Untying the knot releases the energy again: witness the atomic bomb.

This pre-Socratic account of energy as the ur-substance from which everything is made is pervasive in popular science presentations, and is taken with seriousness by some physicists. If you buy this account, then the precise conservation of energy might come to seem sacrosanct. No violations, of any magnitude, are to be tolerated. In a somewhat plausible way, this account of what energy \emph{is}, namely the fundamental substance from which things are made, at least strongly suggests that one ought to expect there to be a perfectly precise conservation law for energy.

\subsection{The pre-Socratic view refuted}

For all the intuitive attraction of the pre-Socratic view, some simple considerations prove that it is nonsense, and ought not motivate placing the conservation of energy in the class of inviolable laws of nature.

When we speak of there being several ``forms of energy'', and say that some energy has ``transformed from one type to another'', there is a standard list of ``forms'' that are invoked: kinetic energy, gravitational potential energy, electric potential energy, thermal energy, chemical potential energy, etc. (this list is in some ways redundant). Top of the list, and the most seemingly palpable and immediately accessible form of energy is the kinetic energy, the energy of motion, $\frac{1}{2} mv^2$. To get hit in the head by a fastball is to become directly and intimately acquainted with some kinetic energy.

But the kinetic energy is the form of energy that completely invalidates the pre-Socratic view. Consider: two icy asteroids of mass $m$ traveling inertially collide inelastically and some of the ice on each asteroid melts. It takes energy to melt ice. So where did the energy come from? Where was it before the collision occurred?

Let the relative speed of the two asteroids before the collision be $V$. Then according to the rest frame of asteroid 1, it began with zero kinetic energy and asteroid 2 had $\frac{1}{2} mV^2$. After the collision, both asteroids, now stuck together, move off at $\frac{1}{2} V$, for a total kinetic energy of $m(\frac{1}{2} V)^2 = \frac{1}{4} mV^2$. The remaining $\frac{1}{4} mV^2$ of the initial kinetic energy went to melt the ice, and it all originated on asteroid 2.

But of course according to the rest frame of asteroid 2 the story is the same with the roles reversed: the kinetic energy that melted the ice all came from asteroid 1. And according to the center-of-mass frame, the two asteroids approach from opposite directions, each traveling at $\frac{1}{2} V$ and so each with $\frac{1}{8} mV^2$ kinetic energy, all of which goes into melting the ice. So who is right and who is wrong?

According to Newton, there would be a right and a wrong here: each asteroid would have an objective and unique Absolute velocity, and its objective and unique kinetic energy would be a function of that. The correct story about energy flow is the story told in the frame of absolute rest.

But we no longer believe Newton. Absolute velocities got thrown under the bus long ago, and absolute objective kinetic energies got thrown out with them. The ``energy that melted the ice'' did not come from anywhere in particular. Even more directly, just as we say that there is no such thing as Absolute velocity, we ought to say that there just is no such thing as kinetic energy or---one would think---any other sort of energy. The whole notion of energy as a substance is a fiction, a holdover from an earlier, and now obsolete, physics\footnote{
This is not meant to imply that
  in each frame there is a problem with energy conservation. As is well known, if we consider a
 system of bodies mutually interacting trough conservative forces, but subject to no external
 forces, then the total energy is conserved with respect to all Galilean frames. The point is that,
 for any specific situation, the values of those energies differ from frame to frame, so energy
cannot be regarded as a sort of
“substance” which might, for instance, be identifiable with what in modern parlance is designated as a “local beable”.}. 

The fall of the pre-Socratic view could not be more stunning. Energy has gone from being the primary substance of which everything is made, completely ungenerable and indestructible, to an outdated fairy story. But that makes our main topic even more pressing: if there is no such thing as energy at all, \emph{then why is it conserved}? We have gone from a view of what energy is that makes the perfectly precise conservation of energy seem inevitable to one that makes even the approximate conservation of energy seem miraculous. 

Fortunately, there is yet a third view.

\subsection{Noether's theorem and symmetry}

An entirely different way to understand what energy---and linear momentum and angular momentum---are was provided by Emmy Noether in 1915. Noether proved that every differentiable symmetry of the action of a system whose dynamics is governed by a Lagrangian generates a conserved quantity for that system. A translational symmetry in a direction generates the quantity we call ``linear momentum'' in that direction; a rotational symmetry about an axis generates the ``angular momentum'' about that axis, and a translational symmetry in time generates a conserved ``energy''. This approach bears no resemblance to the pre-Socratic approach to energy (or to, e.g., the linear momentum as a sort of physical magnitude that a body ``has'' and can ``pass on'' to a body it collides with, like passing a torch in a relay race). Rather, the conserved quantities are, in an obvious sense, not real physical magnitudes or quantities or substances at all. They are merely mathematical shadows of global symmetry properties of the Lagrangian. Despite associating these conserved numbers with individual bodies, as we are wont to do, they arise and have their utility from physical facts that vastly outrun the state of the bodies themselves. The ontological perspective is rather unsettling. We are used to think of the energy as something physical there---in the object. Or even, as the pre-Socratic view has it, \emph{as} the object. But in this account the conserved number arises from quite different sources than we imagined. Nonetheless, the view has many virtues of note.

One virtue is that according to this view, there are as many different ``energies'' as there are different time-like symmetries. If we think of energy as substance-like, then an infinite number of distinct energy-substances is clearly an infinite number too many: we would not know what to do with them all. But we already know that in Galilean space-time and in Minkowski space-time there is an infinitude of distinct global time-like symmetries: one for each distinct global inertial frame.

This resolves the ``Where was the energy?'' puzzle without residue. \emph{All} of the energies are equally real (as conserved quantities arising from a symmetry) and equally unreal (none of them are localized physical magnitudes). Of course, an observer in inertial motion might choose to use the conserved quantity associated with her own inertial frame to do calculations and make predictions, although she could use any other frame's conserved quantity just as well. Each judgment about ``where the energy came from'' is just as valid as the next, as all are about different ``energies''.

In the context of field theories, Noether's theorem associates differentiable symmetries to conserved currents, i.e., currents that satisfy a continuity equation. In particular, the presence in space-time of a time-like Killing vector field $\xi^a $ implies the existence of a conserved current given by $T_{ab} \xi^a $, with $T_{ab}$ the energy-momentum tensor of the matter fields. The continuity equation satisfied by such a current implies that the quantity $\int_{\sigma} dV T_{ab}\xi^a n^b $, with $n^a$ a unit normal to a spatial region $\sigma$, can be interpreted as the ``total energy'' within that region (associated with the Killing field $\xi^a $)\footnote{Note that the quantity $T_{ab}\xi^a n^b$ can only be interpreted as the energy density of some observer if the region in question is orthogonal to $\xi^a$ and if such Killing field can be considered a four velocity (for which it needs to be such that it could be normalized everywhere).}. This is so because, when considering two such spatial regions, the difference in this energy between the two is accounted for by the integral of a flux (expressible in terms of $T_{ab}$ and $\xi^a $) over the boundary of the space-time region interpolating between them. In particular, when the energy-momentum tensor vanishes (or drops to zero in a suitable manner) outside of a compact region, the integral can be taken over the whole space and the total energy becomes independent of time. This means that, under the above conditions, the quantity $\int_{\Sigma} dV T_{ab}\xi^a n^b $ can be shown to be the same for any Cauchy hypersurface $\Sigma$. It is important to stress, this whole construction works for any time-like Killing vector field. Therefore, we will have as many ways of assigning a total energy to a given region as there are time-like Killing vector fields. 

The infinitude of conserved energies constructed via Noether's theorem suffers a startling reversal as soon as Special Relativity is superseded by General Relativity\footnote{For deeper discussions about energy conservation in general relativity see for instance \cite{wald,bondi,curiel}.}. There, in the generic case and certainly for the actual universe, instead of an infinitude of global time-like Killing fields, there are none. There is therefore no reason---if this is the correct account of the nature of ``energy''---to expect any principle of exact global conservation of energy to obtain. The conclusion is admittedly somewhat disconcerting, but there it is.

Despite the fact that, generically, there are no global time-like Killing fields in the general relativistic context, the energy-momentum tensor in such a context continues to be divergenceless. However, one might not use that to define a conserved global quantity in the absence of a Killing field. Relativists often make use of a class of idealized models that admit of exactly one global time-like Killing field (such as a Schwarzschild's description of the space-time metric in the solar system). The practical advantages of such models, which have both a uniquely definable global time function and a unique globally conserved energy, are evident\footnote{One should keep in mind that, in those contexts, the conserved energy of, say, a free particle, $E = -p^{\mu} \xi^\nu g_{mu\nu }$, might not be directly related to the energy associated with any local inertial frame, because the time-like Killing field $\xi^\nu $ is not, in general, a vector with unit norm. That leads, starting from the general relativistic description, to the effective concept of potential gravitational energy, and underlies well-known effects such as the gravitational red-shift observed in precision tests such as the Pound Rebka experiment, \cite{PB}.}. But these practical advantages must not blind us to the fact that the models are idealizations, whose fit to the actual world is always only approximate, and may break down quite badly\footnote{The cosmological setting is one in which no notion of conserved energy is available, as can be seen quite directly by considering, say, the energy in the form of radiation. In that case, the energy density is known to scale with the inverse fourth power of the cosmological scale factor while the spatial volume scales like the scale factor to the power three.}.

The bad news for exact global conservation of energy, on this account, is counterbalanced by good news for approximate local conservation. There are no precise Killing fields in the actual world, but there are myriads of local approximate Killing fields at different scales.

At the scale of a lab, for example, space-time can be well approximated by Minkowski space-time, with its infinitude of energies. This will not do at the scale of the entire earth, but there is an appropriate range of scales where the Schwarzschild metric does a good enough job for most purposes, so there will be, approximately, a conserved Schwarzschild energy. And so on.

If Noether's account of the origin of a conserved (or nearly conserved) quantity called ``energy'' is correct, then our attitude toward the conservation of energy should be similar to our attitude toward Euclidean geometry after the acceptance of General Relativity. In General Relativity, the entire notion of ``the geometry of space'' becomes fundamentally different: to even have a sharply defined spatial geometry you have to specify a space-like surface, and in many cases no particular exact space-like surface will present itself. Having done so, we find that in the actual world, the ``geometry of space'' is never exactly Euclidean. But we also understand why Euclidean geometry worked so extraordinarily well over the whole course of human history, given the precision of measurement that was required. And we also understand that in going up to cosmological scale, there will be no useful role for Euclidean geometry at all. And all of this despite the fact that Euclidean geometry was regarded as \emph{a priori} proven knowledge that could be relied on forever and was perfectly correct.

Is that the fate of the conservation of energy?

\section{Conservation laws in quantum mechanics}

In this section we discuss the status of quantum conservation laws, both in non-relativistic and relativistic contexts. After some general observations, we examine how different alternatives to the standard framework handle the conservation issue and explore implications for the semiclassical gravity program. 

\subsection{The non-relativistic context}
\label{nrq}

As we saw above, the space-time structure on which a physical theory is defined, constrains the type of conservation laws that such a theory can satisfy. However, a theory might not take full advantage of what the corresponding space-time structure has to offer. Think for example of parity violation in the weak interaction: the space-time structure of the theory is symmetric under the interchange of left and right, but the dynamics of the theory is not. In such cases, a conservation law is not satisfied, even though the space-time on which the theory is defined allows for it to be. Non-relativistic quantum mechanics seems to be a case in point.

In quantum mechanics, Hermitian operators that commute with the Hamiltonian are often called \emph{constants of the motion}. However, all this implies is that, during unitary evolution, the expectation values and Born probabilities of such operators are conserved. This suggests that these quantities are conserved, if at all, only \emph{on average} and not on a case-by-case basis. Moreover, \emph{measurements} generically do change expectation values.

To explore the issue in more detail, consider performing an energy measurement on a system $S$ in the state 
\begin{equation}
|\psi\rangle_S = \frac{1}{\sqrt{2}} \left( |E_1\rangle_S + |E_2\rangle_S \right) ,
\end{equation}
with $|E_1\rangle_S$ and $|E_2\rangle_S$ energy eigenstates of $S$ with $E_1 \neq E_2$. Initially, the energy expectation value is given by $\frac{1}{2} \left( E_1+E_2 \right) $, but after the measurement it is either $E_1$ or $E_2$. That is, with the measurement, the energy expectation value jumps, and as a result of the indeterministic nature of standard quantum theory, it sometimes jumps to $E_1$ and sometimes to $E_2$.

Is indeterminism, in and of itself, in tension with a conservation law? Not really. One could perfectly well construct an indeterministic theory in which, say, energy is conserved. Here, though, the indeterminism involved seems to be directly associated with the very quantity for which the conservation is at stake, which seems problematic. So, does the sudden change of expectation value implies that energy is not conserved during the measurement? It is hard to say because the system evolves from an initial state in which energy does not have a well-defined value to a final one that does. It all seems to hang, then, on how the notion of the initial energy not being well-defined is interpreted. If one adheres to the so-called Eigenvector-Eigenvalue (EE) link,\footnote{Such a rule holds that a physical system possesses the value $\alpha$ for a property represented by the operator $O$ if and only if the quantum state assigned to the system is an eigenstate of $O$ with eigenvalue $\alpha$ (see \cite[p. 31]{Alb:92}.} then the non-definiteness of energy must be taken at face value. If so, it is not clear how to evaluate the conservation issue: energy does not seem to be conserved, but it also does not seem to be \emph{explicitly} not conserved. Alternatively, one could try to hold that the system always has a well-defined value of energy, even if the state is not an energy eigenstate. However, such a position would involve some sort of hidden variable theory, which goes beyond the standard formalism. Below we will explore this issue in detail in a well-developed hidden variable theory, namely, the pilot-wave theory, as well as within other alternatives to the standard formalism; for now we continue exploring the conservation issue in the context of the standard framework. 

We can circumvent the non-definitiveness of energy issue discussed above by considering a two-step process. First, consider the system $S$ in the energy eigenstate $|E_1\rangle_S$ and measure some property that does not commute with its Hamiltonian. This will necessarily leave $S$ in a superposition of energy eigenstates; suppose, for simplicity, that such a superposition is precisely the state $|\psi\rangle_S$ above. Then, measure the energy of $S$. By doing so, there is a 50\% chance of ending up on $|E_2\rangle_S$, i.e., a different energy eigenstate than the initial one. This time we seem to arrive at an apparently explicit energy non-conservation. Have we shown that energy is not conserved in quantum mechanics? Not so fast. In the description above we have left out the measuring apparatuses, and it may well be that including them restores exact conservation. After all, energy would be expected to be conserved only in a closed system. So, could including the description of the measuring apparatus lead to an exact energy conservation?

To evaluate the issue, one must first ask if it is in principle possible to give a quantum description of the measuring apparatuses or not. If the answer is no, as Bohr held, then it seems that the statement that including the apparatuses restores energy conservation cannot be explicitly evaluated and can only be assumed, not proved. That is, under these circumstances, it seems is impossible to analyze things, such as energy fluxes between system and apparatus, with any level of rigor and precision. On the other hand, if, at least in principle, it is possible to give a quantum description of the measuring apparatuses, then, in order to have energy conservation, what would be needed is for the state of the closed system, consisting of $S$, the measuring apparatuses, and even the environment, to have the same energy at all stages of the two-step process described above. This however, would be a problem because such a state would then have to be an eigenstate of the total Hamiltonian and, as such, it would no be able to evolve in time---contradicting the fact that stuff, such as measurements, presumably happened during the process. Conversely, if the total state is not an eigenstate of the total Hamiltonian, then there is non-trivial evolution, but the total energy is never well-defined, so it cannot be said that energy is conserved.

To escape this rigid conclusion, one could allow for a small energy uncertainty in the initial state, enough to permit a non-trivial evolution, and ask if the energy of the system will ever wander outside of the energy range allowed by such an uncertainty. Clearly, during unitary evolution, or even in the course of an energy measurement, this will not occur. What happens, though, during measurements of quantities that do not commute with the Hamiltonian? To rigorously answer this question, it is necessary to specify exactly how quantum collapses work. However, the vagueness of the standard formalism in connection with measurements gets in the way, and does not allow for an unambiguous analysis of the issue. One option is to assume, with von Neumann, that the collapse is a global process, in the sense that affects the state, not only of the system to be measured, but also of the measuring apparatus. If that is the case, then measurements of a quantity that does not commute with the Hamiltonian can perfectly well bring the whole system, $S$ plus apparatuses, to states that contain energy components outside of the range allowed by the initial uncertainty. (To see this, consider a state with a small energy uncertainty centered around $E_1$, with $E_2$ outside of the range allowed by the uncertainty, and consider the measurement of an operator with $|E_1 \rangle + |E_2 \rangle $ as an eigenstate.) It is not clear, though, that such measurements can actually be performed in practice. If, on the other hand, collapses somehow only affect the system to be measured, and not the global state, then it is not clear what it is exactly that happens during a measurement and, again, the energy conservation issue cannot be explicitly evaluated.

Another quite interesting system to examine the conservation issue is an EPR-type situation consisting of two subsystems, $A$ and $B$, in the state
\begin{equation}
|\psi\rangle_{AB} = \frac{1}{\sqrt{2}} \left( |E_1 \rangle_A |E_2\rangle_B + |E_2 \rangle_A |E_1\rangle_B \right), 
\label{sup}
\end{equation}
with $ |E_i \rangle_X $ an energy eigenstate of system $X$ with energy $E_i$ (again, with $E_1 \neq E_2$). Note that, in such a state, neither $A$ nor $B$ has a well-defined energy value, but that the energy of the whole system is well-defined to be $ E_1+E_2 $.

Now, imagine $A$ and $B$ initially at the origin, with $A$ moving to the right and $B$ to the left. What happens if, after the systems have separated, we measure the energy of, say, $A$? Before the measurement, $A$ does not possess a well-defined energy value and its energy expectation value is $ \frac{1}{2} \left( E_1+E_2 \right)$. Afterwards, its energy is either $E_1$ or $E_2$. The energy expectation value again jumps from $\frac{1}{2} \left( E_1+E_2 \right) $ to either $E_1$ or $E_2$. In order to explain this jump, one could try to argue that the apparatus that measures $A$ is somehow involved in the balance of energy that ensures energy conservation. Note, however, that the measurement of $A$ also induces a jump in the energy of $B$, but without it interacting with a measuring apparatus. It seems, then, that if measuring apparatuses help resolve an apparent violation of energy, they could do so only non-locally. 

Note however that, throughout this last discussion, we have been acting as if, in this scenario, one can talk of such and such energy being located in such and such a region. However, in this quantum context, it is not clear  that is the case. Think for example of the state $ |\psi \rangle_{AB} $ above: it describes a situation in which the total energy has a well-defined value of $E_1+E_2$, but in which it seems impossible to specify where such an energy is located. Above we saw that, within a classical field theory, we can define the total energy of a spatial region. Below we explore this issue in the context of a quantum field theory.

To sum up, the ambiguous nature of the standard formalism does not allow for a definite answer to the conservation issue. It seems hard to explicitly show that energy is not conserved, but it is equally hard to show that it is. It is not clear, though, that this lack of explicit conservation has any practical implications. Moreover, the non-local aspects of the theory could potentially translate into a sort of non-local behavior for the energy, although this assumes that energy could be thought of as being localized, which seems to be difficult in this scenario. What seems clear is that all of these issues do not lead to inconsistencies or problems of that sort. In the next section we will explore if this is still the case in relativistic contexts.

\subsection{The relativistic context}
\label{qrel}

In non-relativistic quantum physics we found hints of a non-local behavior for the energy of a system. We noticed, though, that in such a context energy cannot really be thought of as localized to a region, in which case, it is not clear what to make of its alleged non-local behavior. Here we explore such an issue in more detail.

Generally speaking, regarding the connection between locality and conservation we seem to have two possibilities. The first option is to have a situation in which energy is localizable to a region, and in which the conservation of energy is a principle that demands that the total energy of the universe (the global integral of this local quantity) not change. In a non-relativistic setting with a unique foliation (absolute simultaneity), it is in principle possible to have global conservation of a local quantity while denying the existence of a local conservation law. In such a scenario, one might be able to associate to localized susbsystems a ``local version of the quantity'' in a manner where it might just disappear from a region without being continuously transported through its boundaries to other regions, but with some kind of global conservation law ensuring that it simultaneously appears somewhere else. That would mean one would have global conservation of a local quantity without local conservation, and some sort of ``discontinuous transmission''. Note however that for such scenario to be viable it is essential for there to be a local quantity, something like an ``energy density'', even if it is not accompanied by another quantity corresponding to an energy flux (such as the one we described in the context of classical field theory), so that one might keep ``track'' and identify at each instant where it is. That is, one should be able to refer to ``the energy'' present in a given spatial region in a given leaf of the preferential foliation as the integral of some sort of energy density over that region\footnote{It is in principle logically possible for such an assignment to be made to broadly defined regions, without there being an actual notion of energy density. It seem however very hard to imagine how such a recipe could be framed precisely and in general without such a concept.}.

This form of global-conservation-without-local-conservation would be possible in a setting where the theory provides a preferred foliation, but not in a case in which one is free to use whatever foliation into space-like hypersurfaces one likes. In such a case, it is easy to run the foliation through the space-like gaps between the disappearance of the local quantity in one location and its reappearance elsewhere. In such a foliation one will either have double counting of the local quantity (it will appear twice on the same hypersurface) or else a leaf of the foliation on which it goes missing altogether. In either case, the global conservation law will fail. So global-but-not-local conservation of a local quantity is possible in settings with a preferred foliation but it would not be possible in relativistic contexts where the choice of foliation is open.

Alternatively, energy could be something not constructed out of a local quantity such as energy density\footnote{The assignment of the so-called ADM mass to asymptotically flat space-times provides a notion of energy with these characteristics: it is a quantity that is computed as a certain infinite size limit of a surface integral that \emph{cannot}, in any straightforward way, be expressed as an integral over a space-like Cauchy hypersurface in that space-time (see for instance \cite{HawkingEllis}).}. If this is the case, then the previous argument does not go through. If energy is not the integral of a local quantity, then one cannot identify ``the region from which the energy disappears'' and the ``region at which the energy reappears'' and run a leaf of the foliation through the gap between them. So if there is to be an argument that energy cannot be globally conserved in such a case it has to take a different form. The point is that in such a situation one would not be able to define an energy density that behaves appropriately.

Above we saw that, within classical field theory in general, and within special and general relativity in particular, presence of time-like symmetries allows one to define a conserved energy-momentum tensor out of which energy might be defined in association with a region of space. On the other hand, we saw that in standard, non-relativistic quantum mechanics, energy generically cannot be thought of as a quantity localized within a region. What happens in this regard in the context of quantum field theory?

In quantum field theory, one works with the energy-momentum tensor operator $ \hat{T}_{ab} $. In the presence of a time-like Killing field $\xi^a$, by integrating $ \hat{T}_{ab}\xi^a n^b $ over a region $\sigma$ (with $n^b $ its normal vector) one can construct an energy operator associated with such a region. This, however, does not mean that, generically, energy can be ascribed to particular regions of space because, generically, the state of the field is not an eigenstate of such an operator. One can consider the expectation value of these energy operators and, if their uncertainty is small, take the expectation value as approximately representing the energy content of the corresponding region. But, is this identification of the expectation value with the actual value reasonable? And what happens if the uncertainty is not small? This questions take us back to key conceptual issues within the standard interpretation which, simply put, do not have good answers. In particular, the lack of ontological clarity, i.e., of a clear account of what the theory actually talks about, gets in the way of a proper analysis of the issue. Bellow we will be able to return to this matter and explore it in more detail in the context of alternatives to the standard framework that correct this shortcoming.

Leaving behind the non-locality issue for now, what can we say about global energy conservation in quantum field theory? In the context of Minkowski space-time, one could argue that the symmetries of the situation, together with Noether's theorem, guarantee the fulfillment of space-time-related conservation laws. However, what Noether's theorem implies in this context are conservation equations in terms of operators and, as we saw in the non-relativistic quantum context, such equations do not directly translate into actual conserved quantities. That is, the complications we encountered in the non-relativistic setting regarding, for example, the fact that energy, generically, does not have a well-defined value, even if its expectation value is conserved, persist in this setting. Moreover, we are leaving out of the picture the collapse process, which complicates matters enormously---both regarding conservation laws and the viability of the standard framework.

What about quantum field theory on curved space-times? In such a framework, on top of the issues we encountered in the context of Minkowski, we have the fact that space-time, generically, does not posses the symmetries required for conserved Noether currents to arise. It is clear, then, that conservation laws are completely absent in such a setting. What to make of this? Well, even if it is the case that conservation laws are not satisfied in a relativistic quantum field theory, that would not be such a serious problem. Of course, such laws, when available, remain very useful in practice, but they are not essential, as the self-consistency of quantum field theory does not depend on their fulfillment. On top of all this, as in the non-relativistic case, it is not even clear in this setting that it would be possible to construct the macroscopic superpositions that would render the non-conservation situations relevant \emph{in practice}.

Much more interesting would be to explore the status of conservation laws in a general relativistic, quantum context---in which, unlike in quantum field theory on a fixed background, \emph{backreaction} effects are incorporated. The obvious problem, however, is that we do not yet have at our disposal a complete theory of quantum gravity to do so. Still, even in the absence of such a framework, there are a few thing we can say.

The quantum nature of matter implies the necessity to construct a theory of gravity that, unlike general relativity, acknowledges such a quantum character\footnote{It is often assumed that constructing a theory of quantum gravity is synonymous with quantizing the metric itself. However, this is not necessarily the case. In fact, we would not really know how to make sense of the notion of a superposition of different metrics. Note that this is in contrast with the much easier task of considering, say, a superposition of electromagnetic fields. In that case, we could, for instance, point to a certain region of space-time and argue that the value of the electric or magnetic field we would obtain if probing such a region with a test charge will be so and so, with probabilities this and that. In the case of the superposition of two space-time metrics, on the other hand, we would not even be able to point to any region to argue that our wave function indicates what we are likely to obtain, if such region were to be probed with test particles. We simply would not be able to identify, in the various metric superpositions, which subsets of the manifolds would correspond to the ``same'' spatiotemporal physical region.The implicit tension described above is illustrated by Penrose's time uncertainty argument \cite{Penrose-Emperors}. There he argues that such situation is only likely to be resolved if one assumes that somehow physics prevents the persistence of such quantum superposition of widely different space-times. That, in turn, is used to argue in favor of a quantum gravity induced collapse of the wave function.}. A very likely implication of this is that classical space-time will become no more than an effective phenomenon that, at some level of approximation, emerges out of the fundamental, quantum gravity degrees of freedom. In fact, most approaches to quantum gravity do not pose spatio-temporal degrees of freedom at the fundamental level\footnote{Or at least not standard ones. For instance, Loop Quantum Gravity relies on degrees of freedom that have purely spatial character, while the Causal Sets approach relies on degrees of freedom that are binary and represent just causal relations, forsaking any metric notions at the level of the fundamental entities.}. Still, in order for them to be empirically viable, they are required to explain how, at some level of approximation, an essentially classical space-time emerges\footnote{The problem of recovering space-time notions from full quantum gravity proposals is incredibly difficult. For instance, canonical approaches to quantum gravity lead to timeless theories in which the recovery of space-time notions is possible only in an effective level and after a series of approximations and manipulations (that are as far from being under-control as the proposed theories of quantum gravity themselves).}. That is, even if at the fundamental level such theories have nothing resembling a classical space-time, they are required to make contact with the semiclassical scenarios, containing an emergent space-time, in which they could be experimentally probed. As a result of all this, regardless of the ultimate details of quantum gravity, such a theory must, at some level, be well-approximated by a semiclassical scenario, in which matter fields are treated quantum mechanically, but space-time is treated classically. 

Regarding the status of conservation laws in quantum gravity, we note that the notion of conservation, by its very meaning, needs to be framed in a context where space-time must be taken for granted. Therefore, an analysis of the conservation issue in this context must be carried out in a regime in which an essentially classical description of space-time is already appropriate. In any case, exploring the semiclassical regime is probably the best we can hope to do with reasonable rigor, given the fact that we do not yet have a complete theory of quantum gravity.

The basic equation of the semiclassical formalism is given by
\begin{equation}
G_{a b} = 8 \pi G \bra \psi \hat T_{a b } \ket \psi ,
\end{equation}
in which the classical Einstein tensor is coupled, not to the classical energy-momentum tensor, but to the expectation value of the energy-momentum tensor of the quantum matter fields (with $\ket \psi$ its quantum state). We note that, as a result of the Bianchi identities, the divergence of the left-hand-side of the equation is always zero\footnote{We note that the Bianchi identities depends crucially on the smoothness of the metric tensor and it is unclear if a less differentiable object, say a metric that is only continuous or just $ C^{1}$ would not fall to obey it.}. However, there is nothing within quantum theory that guarantees that the divergence of the expectation value on the right-hand-side will also (generically) vanish---particularly during quantum collapses. Does this constitute a fatal blow for the semiclassical program?\footnote{It is worth noting that in a certain variant of Einstein's general relativity known as \emph{unimodular gravity}, the conservation of the energy-momentum tensor is not realy a requirement for self-consistency. Such a feature was used in \cite{JosPerSud:17,PerSudBjo:18} as the basis for a proposal to address the nature and actual value of the so-called dark energy component of the universe.}

The self-consistency of semiclassical gravity has been called into question repeatedly over the years. For instance, in \cite{EpplyHannah:1977} it is argued against semiclassical gravity with a thought experiment in which the position of a quantum particle, assumed to have small uncertainty in momentum, is measured with a classical gravitational wave with small wavelength \emph{and} small momentum. According to the authors, there are three different possible outcomes of the experiment: i) the measurement localizes the quantum particle to within the gravitational wave without transferring a large momentum---in which case Heisenberg's uncertainty principle is violated; ii) the measurement localizes the quantum particle without violating Heisenberg's uncertainty principle---in which case energy conservation is violated; iii) the measurement does not collapse the wavefunction of the particle---in which case the measurement could be used for superluminal communication. It turns out, however, that in \cite{HugCal:01} it is explained why the argument in \cite{EpplyHannah:1977} falls short of a no-go theorem. Worse yet, in \cite{Mat1,Mat2} it is shown that the proposed experiment cannot be carried-out, even in principle.

The viability and experimental adequacy of semiclassical gravity are also considered in \cite{PagGei:81}, in which an actual experiment is reported. The experimental set-up employed is one where a quantum event (e.g., the spin along $z$ of a spin-$\frac{1}{2}$ particle prepared spin-up along $x$) is used to determine whether a massive sphere $M$ is displaced towards the right or left\footnote{The actual experiment involved a more complex scenario, but those details are not relevant for the present discussion.}. This is followed by the measurement of the gravitational field of such an object through a Cavendish-like torsion balance using a small test mass. Needless to say, the result is what one would expect: when the quantum event results in a displacement towards one side, the Cavendish test mass experiences a gravitational field at the corresponding location.

This rather unsurprising result has been greeted with great enthusiasm by some advocates of the quantum gravity program, and viewed in some quarters as an indication of the non-viability of semiclassical gravity. The basic argument is the following: if we take the point of view that there is no quantum collapse of the state vector, then the spin measurement will leave the spin-$\frac{1}{2}$ particle in a superposition of spin-up and spin-down along $z$. Under such conditions, the mass $M$ will get into a superposition of being displaced toward the right and towards the left and the gravitational effect of such a mass will retain the symmetry of the initial condition, ensuring that the test mass will not move either way. This, however, is contradicted by the experimental result. On the other hand, if the quantum state of the system does undergo some sort of collapse, then the divergence of $\bra \psi \hat T_{a b } \ket \psi$ during such process will not (generically) vanish and the semiclassical equation will be violated. 

This seems like a strong argument against semiclassical gravity. However, we do not take it to be conclusive (see also \cite{Carlip}). To see this, it is important to notice that \cite{PagGei:81} highlights two potential problems with the semiclassical framework. On the one hand, it reminds us of the fact that during quantum collapses the divergence of the expectation value of the energy-momentum tensor might not vanish, as is necessary to ensure self-consistency. On the one hand, it shows that quantum states with a large dispersion in $\hat T_{a b } $, such as that of a macroscopic object in a superposition of different positions, may render the theory \emph{empirically} problematic.

Regarding the first issue, we first note that the argument relies upon a rather simplistic account of the measuring process present in the standard formalism. Below we will explore the issue within several alternatives. At any rate, we acknowledge that the issue constitutes a serious hurdle for taking semiclassical gravity as a \emph{fundamental} description of the situation at hand. However, we do not see it as an impediment for looking at it as a viable \emph{effective} description, with restricted but rather wide applicability. An analogy is provided by the standard reading of the Navier-Stokes equations as applied to fluid dynamics. Such equations provide a robust description of fluids under a wide set of circumstances. Yet, they represent no more than a good approximation, that does not contain the fundamental degrees of freedom and breaks down under certain circumstances (i.e., turbulence or the break of a wave in the ocean). Similarly, we can expect the semiclassical description of space-time to breakdown in some situations, particularly in connection with what in the standard approach would be associated with the collapse process.

Regarding the second issue, a key question is to what extent it is in fact possible to construct quantum states with large enough energy dispersions to make semiclassical gravity empirically inadequate. For example, it might be argued that at the practical level, because of decoherence, the superposition of $M$ being at two distinct places considered in \cite{PagGei:81} never actually realizes. The problem, though, is that, once more, the vagueness of the standard formalism does not really allow for a general, rigorous assessment of the issue. That is, the ambiguities associated with the collapse postulate in standard quantum mechanics get in the way of deciding if it is actually possible \emph{in practice} to prepare a macroscopic system in a quantum superposition of very different energy eigenstates. In the next section we will explore in detail what theories that improve upon this vagueness of the standard framework imply regarding the construction of such macroscopic superpositions. In order to do so, we introduce a scheme that, based on the fact that in general relativistic contexts there is simply no general law of energy conservation, allows for the construction of quantum superposition involving arbitrarily large energy gaps. The \textbf{gedankenexperiment} we are proposing goes as follows:
\begin{quote}
Start with a photon with energy $E$ and make it pass through a half-silvered mirror $M_1$, such that one path takes the photon to a box\footnote{To be precise, we envision a cavity surrounded by perfectly reflecting mirrors, massive enough to ensure that the photon energy loss, as a result of the multiple recoils over the whole experiment, is negligible. Although this might be quite difficult to achieve in practice, there is no known feature of standard quantum theory indicating this to be unfeasible in principle.} and the other sends it to a second mirror, $M_2$, placed on a far away galaxy, receding from us due to the cosmological expansion. The photon traveling to $M_2$ will arrive back to earth with energy $ \chi \chi' E $, with $\chi$ and $\chi'$ the to and fro red-shift factors. If we now direct such a photon to the box, we obtain a photon in the state $\frac{1}{\sqrt 2}(\ket E + \ket{\chi \chi' E}) $. Next we can consider a collection of $N$ entangled photons being sent through a shutter that, depending on a quantum decision (e.g., the spin along $z$ of a spin-$\frac{1}{2}$ particle prepared spin-up along $x$), lets them either go directly to the box or travel first to $M_2$ and then go to the box. If so, we end up with a superposition of two energy eigenstates: one with energy $NE$ and one with energy $ N \chi \chi' E$. By letting $N$ grow, the energy difference can be made arbitrarily large.
\end{quote}
The idea, again, is to use the gedankenexperiment as a testing ground to analyze the extent to which different quantum frameworks allow for superpositions of energy states that could jeopardize the empirical viability of semiclassical gravity. The objective is to explore if it is possible to construct a quantum framework with which semiclassical gravity yields a good approximation where it actually should, that is, where general relativity has made predictions confirmed to within experimental tolerance. The game, then, is to make sure that the deviations from general relativity are small, but only where they ought to be.

Many of the issues that led to trouble regarding a rigorous analysis of the status of conservation laws in quantum contexts appear to be intimately connected with the conceptual problems of the standard framework. It is clear, then, that to continue this analysis we can no longer ignore them. In what follows, we analyze the status of conservation laws in three of the most promising alternatives to the standard formalism, namely, objective collapse models, the pilot-wave theory and Everettian frameworks. We do so in order to determine the specific form the difficulty takes in each one of them, and the extent to which they offer possible paths to eliminate the problem or, at least, to reduce the impact of the expected violations.

\subsection{Conservation laws in alternative quantum frameworks}
We saw above that the weaknesses of standard quantum mechanics, both with respect to its lack of ontological clarity and the vagueness with which it treats the measuring process, impede a rigorous analysis of the conservation issue. Many strategies have been suggested in order to deal with these conceptual problems of the standard framework, but only three of them stand out as serious contenders, \cite{Mau:95}: Everettian proposals that try to do without the collapse postulate \cite{Eve:57}, hidden variable theories, such as the pilot-wave theory \cite{Bohm}, that do not take the wave function as providing a complete description of a system, and objective collapse models, that replace the standard collapse postulate with objective, spontaneous collapses \cite{Bassi}. Our aim below will be to explore the fate of conservation laws within the context of these alternatives.

\subsubsection{Objective collapse models}
By adding non-linear, stochastic terms to the Schr\"odinger equation, objective collapse models seek to provide a single evolution equation that adequately incorporates both the unitary evolution and the standard collapse process. The idea is to construct a formalism that, at the micro-scale, is empirically indistinguishable from standard quantum mechanics, but in which embarrassing macroscopic superpositions, such as Schr\"odinger-cat states, are effectively suppressed---all this without the need of invoking measurements or observers.

The simplest non-relativistic collapse model, known as GRW \cite{GRW}, proposes that all elementary particles randomly undergo spontaneous localization events around positions selected according to a probability distribution that resembles the Born rule. In order to closely approximate standard quantum mechanics at the micro-scale, the spontaneous localization frequency for each particle is postulated to be extremely small (e.g., $10^{-16}$ s$^{-1}$). However, such frequency is automatically enhanced by the dynamics by increasing the number of particles involved, leading to a large collapse frequency for macroscopic objects. This, together with the fact that when a macroscopic object is placed in a superposition of different positions, the localization of any of its constituents yields the localization of the whole object, leads to the desired effective suppression of macroscopic Schr\"odinger-cat states.

The Continuous Spontaneous Localization (CSL) model \cite{CSL} replaces the discontinuous GRW jumps with a continuous, stochastic and nonunitary evolution. In particular, the model modifies the Schrödinger equation with the addition of non-linear, stochastic terms designed to randomly drive any initial wave function towards one of the eigenstates of a so-called collapse operator, with probabilities given by the Born rule. The collapse operator is normally chosen in association with position or mass density because such options lead to a suppression of superpositions of macroscopic objects at different locations, and thus to a solution of the measurement problem.

A possible worry regarding the viability of collapse models arises from the fact that they do not lead systems to eigenstates of the collapse operator, only to states that are, in a sense ``very close'' to them. Therefore, if, as in the standard interpretation, one subscribes to the EE rule, systems under collapse dynamics never actually end up possessing well-defined values for the property associated with the collapse operator---nor for most others. It seems then that, for this models to work, one cannot \emph{interpret} them in terms of EE and a different interpretation---that specifies the relation between the mathematical and the physical objects---is required.

One proposal to solve this issue involves the use of the so-called \emph{fuzzy link}, in which one allows for some tolerance away from an eigenstate to ascribe the possession of well-defined properties. It not clear, though, how to define in a non-arbitrary way how close a state needs to be to an eigenstate in order for the value of a property to be well-defined, \cite{fuzzy}. Moreover, this type of interpretation remains as ontologically obscure as the standard interpretation because it only talks about possession of properties but remains silent regarding what are supposed to be the property bearers of the theory.

A more attractive alternative is to construct out of the wave function a so-called \emph{primitive ontology}, which is interpreted as the stuff that populates the world. The most popular options in this direction are the flash ontology, in which the centers of the GRW collapses are taken to constitute the primitive ontology, and the mass density ontology---available both for GRW and CSL---in which a mass density in 3D space is constructed out of the wave function as the expectation value of the mass density operator.

As is well-known, the energy expectation value is \emph{not} conserved within objective collapse models. For example, when a spread-out wave function undergoes a GRW-type collapse, it gets localized to a smaller region and its energy expectation value rises. In particular, for a free particle of mass $m$ starting in an energy eigenstate, a GRW hit produces an energy expectation value increase of $\alpha\hbar^2/4m$, with $\alpha$ the GRW parameter determining the collapse scale. What are we to make of this non-conservation? In order to answer such a question, it is first necessary to enquirer about the notion of energy within these type of theories, which is not a straightforward matter. As we explained before, the assignment of property values in this theories cannot be performed via the EE rule, so how can the energy of a system be defined?

One option is to employ the fuzzy link to define the energy of a system. However, by doing so, we would not advance much in removing the vagueness that obstructed a clear assessment of the conceptual issues we encountered within the standard context. One could argue that, as long as the fuzzy link provides a sharp-enough energy for each quantum state, one would have a basis for talking about whether it's conserved or not, and if not, how big the violations can get. However, the problem with this perspective is that it is not clear what is meant by ``sharp-enough''. The point is that it is not clear that one can find some universally valid degree of sharpness, and a correspondingly universally valid degree of maximum possible violation, to allow for a rigorous assessment of the conservation issue.

It seems to us that a more interesting option would be to define energy using the primitive ontology and not the wave function, although it is not clear exactly how one could do it. It is interesting to note that, within the mass density framework, a total mass could be defined as the integral of mass density over all space. Since the wave function is always normalized, such a total mass is conserved. However, the discontinuous GRW evolution of the wave function implies a discontinuous evolution for the mass density. Therefore, what we have here is a concrete example of a global-conservation-without-local conservation---which, as we explain above, is only possible in a setting with a preferred foliation, as the non-relativistic one we are considering. 

The original objective collapse models described above are non-relativistic. Recently, though, fully relativistic versions have been developed. For example, in \cite{Tumulka} a relativistic version of GRW for $N$ non-interacting spin-$\frac{1}{2}$ particles in an external field is introduced. In addition, \cite{Bed} develops a Lorentz Invariant, objective collapse model for a relativistic quantum field. To do so, a quantum state is assigned to every Cauchy hypersurface and a stochastic Schwinger-Tomonaga-type equation, that implements the evolution between hypersurfaces, is postulated.

What about possible primitive ontologies in this context? The natural generalization of the mass density ontology to this quantum field theory setting is to introduce as primitive ontology the expectation value of the energy-momentum tensor. However, since the expectation value at a point, which is supposed to represent actual energy-momentum at such a point, is hypersurface-dependent (simply because the quantum state is so), a Lorentz invariant construction would require some covariant way to select the appropriate hypersurface. One possibility is to take the expectation value at a point using the state defined on the past light cone of the point in question (see \cite{RevMat}). It is important to mention that, with this proposal for a primitive ontology, the semiclassical approach becomes much more natural because the expectation value that couples to $G_{ab}$ in the semiclassical equations directly represents the energy and momentum distribution predicted by the theory. Note however that such an expectation value is not conserved (we will return to this issue later).

Above we noticed two potential problems with a semiclassical framework: i) the fact that the divergence of the expectation value of the energy-momentum tensor might not vanish, as is necessary to ensure self-consistency and ii) that quantum states with a large dispersion in $\hat T_{a b} $ may render the theory \emph{empirically} problematic. Could objective collapse models help alleviate this problems?

Regarding the first issue, it is clear that, within objective collapse theories, the divergence of the expectation value of $\hat T_{a b} $ generically does not vanish. As a result, strictly speaking, semiclassical gravity, together with collapse theories, is an inconsistent proposal. Nevertheless, as we mentioned above, we could rescue the proposal by looking at it as an effective description, not expected to hold exactly at all times. That is, just as the hydrodynamic description of a fluid via the Navier-Stokes equations does not hold in some situations (e.g., when a wave is breaking in the ocean), we could take the semiclassical Einstein equations not to hold when objective collapses occur, but to do so before and after. A formalism based on this general approach has been proposed in \cite{P6}. It relies on the notion of a \textit{Semiclassical Self-consistent Configuration} (SSC):
\begin{quote}
The set $\lbrace g_{ab}(x),\hat{\varphi}(x), \hat{\pi}(x), {\cal H}, \vert \xi \rangle \in {\cal H} \rbrace$ is a SSC if and only if $\hat{\varphi}(x)$, $\hat{\pi}(x)$ and $ {\cal H}$ correspond a to quantum field theory for the field $\varphi(x)$, constructed over a space-time with metric $g_{ab}(x)$, and the state $\vert\xi\rangle$ in $ {\cal H}$ is such that
\begin{equation}\label{scc}
G_{ab}[g(x)]=8\pi G\langle\xi\vert \hat{T}_{ab}[g(x),\hat{\varphi}(x)]\vert\xi
\rangle ,
\end{equation}
where $\langle\xi| \hat{T}_{\mu\nu}[g(x),\hat{\varphi}(x)]|\xi\rangle$ stands for the expectation value in the state $\vert\xi\rangle$ of the renormalized energy-momentum tensor of the quantum matter field $\hat{\varphi}(x)$, constructed with the space-time metric $ g_{ab}$. 
\end{quote}
The SSC formulation allows for the incorporation of objective collapses into the semiclassical picture. The idea is for two SSC's to describe the situation before and after the collapse, and for them to be suitably joined at the collapse hypersurface. In detail, the idea is that when a collapse of the matter fields occurs, it leads to a change in the expectation value of the energy-momentum tensor which, in turn, causes a modification of the metric. Note however that all this requires a change in the Hilbert space to which the state belongs, so the construction of a new SSC is required. It is important to mention that this program of adapting objective collapse models to semiclassical setting has shown great promise in the resolution of a list of open problems in cosmology and quantum gravity, such as the emergence of seeds of cosmic structure, the black hole information issue, the problem of time in quantum gravity (see \cite{weight} and references therein).

Of course, something along the lines of this formalism could work if, as in GRW, collapse events are discrete. It is not clear, on the other hand, what to make of the semiclassical framework in conjunction with a collapse model, such as CSL, in which the expectation value of the energy-momentum tensor (almost) never vanishes. The natural path would be to generalize the SSC scheme to the continuum by some sort of limiting process. However, the actual construction of a mathematically rigorous version of such model at this point clearly posses formidable technical difficulties.

What about the second issue, namely the fact that quantum states with a large dispersion in $\hat T_{a b} $ may render semiclassical gravity \emph{empirically} problematic? Do objective collapse models have resources to limit the development of such states? As we explained above, in these theories the \emph{effective} collapse rate of a system is proportional to the number of entangled particles in a superposition of different positions. That is, the main aspect controlling the effective collapse rate is the sheer size of the system subject to a superposition. This feature might seem to open a path for these theories to restrict the generation of the sorts of states that could lead to gross violations of semiclassical gravity. Let us look at this in more detail.

The suppression of superpositions of large objects, present in objective collapse models, could limit the generation of states involving large uncertainties in energy. That is because, as the energy difference in the various energy states involved in the superposition increases, it seems that one would have to increase as well the number of particles involved. However, we must note that the amplification mechanism of the collapse rate in GRW and non-relativistic CSL would not necessarily be at play if the components with different energies are not, at the same time, differentiated with regard to their position. In particular, looking back at the gedankenexperiment, it would seem that the spontaneous collapses would offer only a very slow decay of the superposition of the state of the photons.

There is, however, another essential component of the construction that might be of help, namely, the mirror on which the photons bounce on the far away galaxy. This mirror must be absorbing momentum from the bouncing photons and this recoil momentum would increase as the numbers of photons involved increases. This would lead to a delocalization of the particles that make up the mirror, leading to a state of macroscopic superposition of different positions for that large object. And those superpositions are of course subject to a very rapid localization as a result of the collapse dynamics which will, in turn, as a result of the correlations with the photons in our box, result in a rapid collapse of the state of these photons to one with relatively well-defined energy. How rapid and effective is this process will naturally depend on the total mass of the mirror: as it becomes more massive, its recoil velocity (for a fixed number of bounced photons) will decrease, but the number of particles involved in the delocalized state would increase. These two behaviors would contribute to modify in opposite ways the overall rate of spontaneous collapse, so it is not clear what is going to happen.

At any rate, it is important to note that, at this point in time, collapse theories are very much a work in progress. The explorations so far, mostly done at the non-relativistic level, suggest that the mass density operator must play a central role. In fact the proposals where the collapse rate parameter is proportional to the mass of the particle species in question seems to be phenomenologically preferred \cite{PearleSquires}. This strongly suggests for the collapses operator in the relativistic context to be associated with the energy-momentum tensor (see \cite{Bed,RevMat}). If so, the theory would precisely avoid the large uncertainties in the expectation value of $\hat T_{ab}$ that could lead to predictions of semiclassical gravity in conflict with observations\footnote{It is interesting to note the connection between these ideas and those by R. Penrose \cite{Penrose-Emperors} in which states of sufficiently large gravitational energy dispersion trigger spontaneous collapses.}. We must point out, however, that the different components of the energy-momentum tensor operator do not commute among themselves, so they do not share eigenstates. The best we can aim for, then, is for a collapse theory that drives superpositions to states that minimize the uncertainty associated with the different components of energy-momentum. Is that enough to avoid disagreements between theory and observation? It is not clear at this point.

In sum, objective collapse theories do not appear to have much to offer regarding the enforcement of the divergencelessness of $T_{ab}$, required for self-consistency of the semiclassical equations. However, they seem promising in preventing the occurrence of superpositions of arbitrarily different energy eigenstates. More generally, spontaneous collapse theories seem to significantly expand the space of possibilities for exploring the semiclassical regime. Moreover, they might point to concrete situations where the semiclassical approximation might break---opening the path to novel quantum-gravitational effects (which might not even involve the Planck scale).

\subsubsection{Pilot-wave theory}
Another promising approach to the conceptual problems of standard quantum mechanics is provided by the de Broglie-Bohm pilot-wave theory, which is the best-developed example of a so-called hidden variable theory. The framework was originally formulated in the context of non-relativistic quantum mechanics, where it is assumed that a complete characterization of an $N$-particle system is given by its wave function $ \psi (x, t) $ \emph{together} with the actual positions of the particles $\{X_1 (t),X_2 (t),\dots,X_N (t)\}$, which are taken to always possess well-defined values. The wave function is postulated to satisfy at all times the usual Schr\"odinger equation
 \begin{equation}
 i \hbar \frac{\partial \psi}{ \partial t } = - \sum_k \frac{\hbar^2}{2 m_k} \frac{\partial^2 \psi}{\partial x_k^2} + V(x)\psi 
 \end{equation}
and the positions to evolve according to the so-called guiding equation, 
\begin{equation}
\frac{d X_k(t)}{dt} = \frac{\hbar}{m_k} \left. \text{Im}\left[ \frac{(\frac{\partial \psi}{\partial x_k })}{ \psi} \right] \right\rvert_{x=X(t)}.
\label{guide}
\end{equation}
The pilot wave theory presupposes an original \emph{epistemic} uncertainty about the initial position of the particles, which is taken to be proportional to $|\psi|^2$. Under those conditions, the theory is known to lead to the same statistical predictions as standard quantum mechanics.

As is well-known, hidden variable theories are \emph{contextual}, meaning that results of measurements of certain properties dependent on the details of experimental arrangement, \cite{Bel:66,Koc.Spe:67}. In the pilot-wave theory, in particular, all properties, except the positions of the particles, are contextual. In order to explore this feature in the case of energy, let us consider a particle in a box (represented by an infinite square well potential) prepared in an energy eigenstate. In such a system, the phase of the wave function is a function of time alone and not of space, so the right-hand-side of Eq. (\ref{guide}) vanishes and the particle remains at rest inside the box. Still, if the box were to be opened, the particle would acquire a non-zero momentum and escape. Where does the required kinetic energy come from? From a classical point of view, if the particle is in a higher energy eigenstate then it ought to have a correspondingly higher kinetic energy, and hence be moving faster. However, according to the guidance equation, the particle will be at rest no matter which energy eigenstate it is in. The energy content can therefore not be attributed to an energy carried by the particle.

One might wonder why this does not create empirical problems once the box is opened. But as soon as the potential barrier is removed, the wavefunction starts to evolve by Schrödinger's equation, and that in turn creates a gradient in the complex phase which causes the particle to start moving. And the higher the energy eigenstate the faster it moves. Hence the ``high energy'' particles exit the box quicker and find their way to a distant screen faster, even though all the particles were at rest before the box was opened. It is as if the ``higher energy'' particles were bouncing back and forth faster in the box before it was opened, even though all the particles were at rest. Let us look at all this more generally.

The pilot-wave theory seems to only deal with positions, so it might appear to be limited in scope with respect to the standard formalism, which provides probabilities for all types of observables. The truth, however, is that talk of positions is enough, even within the standard framework. That is because, what we actually do when we measure something, is to read the result of the experiment by looking at the final position of, say, the pointer of an apparatus designed to track the property in question. Consider, as above, a measurement of the energy of a particle in the context of a pilot-wave theory. What is needed to perform such a measurement is for the particle to interact with another system in such a way that, from the post-interaction configuration of some sort of pointer of that system, the energy of the particle can be read. In particular, the different energy eigenstates of the particle must get entangled with the position of the pointer in a way that, by detecting the pointer in a certain position, we can infer which energy eigenstate contains the particle (as in the example above where the particle is liberated from the box, the pointer could be the particle itself).

The important point is that, as soon as the measurement procedure is started, the wave function changes, which causes the trajectory of the particles to change as well. The upshot is that, according to the pilot-wave theory, measurements of properties other than positions do not reveal pre-existing values; instead, results are ``created'' during the measurement. And not only that, such results depend on the details of the experiment.

Given that, as we just saw, energy is contextual in the pilot-wave theory, there seems to be no option for it to be conserved. One could argue, however, that the energy we were talking about above is the energy of the particle, and not that of the whole, particle-plus-pointer, system. Maybe, then, a conserved energy of the complete system could be defined. If so, such a quantity would have to be defined in a non-contextual way. Above we mentioned that, in the pilot-wave theory, only the positions of the particles are non-contextual. However, the precise thing to say is that only the wave function, the positions of the particles, and functions of these two, are non-contextual. As a result, energy could be defined in a non-contextual way if it is written as a function of the wave function, the positions of the particles or both.

For instance, employing only the wave function, one could define energy in a non-contextual way either via the EE link or by defining it as the expectation value of the Hamiltonian. In the first case, energy would only be well-defined if the wave function is an energy eigenstate. In the second case, energy would always be well-defined---and it would even be generically conserved. The problem with these definitions, though, is that they completely ignore the actual positions of the particles, which are regarded by the theory as central elements of its ontology. As a result, any sensible definition of energy in this context must take into account the positions of the particles. We could, then, try to assign an energy using only the positions of the particles via, e.g., $ E = \sum_i \frac{1}{2} m_i \dot{X}_i^2 + V(X) $. This, however, seems rather unattractive because such a quantity is not the one that is conserved during the evolution of the configuration variables.

The most promising alternative seems then to be to define energy in terms of both the positions and the wave function. A natural definition in this regard is given by
\begin{equation}
E = \sum_i \frac{1}{2} m_i \dot{X}_i^2 + V(X) + Q(X)
\label{EQ}
\end{equation}
with $Q(x)$ the so-called quantum potential, given by
\begin{equation}
Q = \sum_i - \hbar^2 \frac{\nabla_i^2 |\psi| }{2 m_i |\psi|} .
\end{equation}
The energy above defined satisfies
\begin{equation}
\frac{\text{d} E}{\text{d} t} = \frac{\partial}{\partial t} \left(Q + V \right) | _{x=X(t)},
\end{equation}
so it is conserved if both $V$ and $Q$ are time-independent. The problem, however, is that $Q$ does not need to be time-independent, even when $V$ is. As a result, the classical conditions for an energy conserving motion do not imply a quantum mechanical conservation of energy. That is, the notion of energy so defined is generically not conserved, even in situations in which one would expect it to be (e.g., when $V=0$).

Another interesting mixed option is given by
\begin{equation}\label{psiHpsi}
E = \left. \frac {\psi^* H \psi }{\psi^* \psi} \right\rvert_{x=X(t)}.
\end{equation}
With this definition, when the wave function happens to be in an energy eigenstate, the result is just the corresponding eigenvalue---and it is conserved. Moreover, if the wave function is a superposition of different energy eigenstates, and such eigenstates have a disjoint support in configuration space, energy behaves as if only the branch occupied by the pilot-wave particles exists. The problem, though, is that when the above conditions are not satisfied, this quantity might not even be real\footnote{One should avoid being confused here by the standard argument that a quantity such as
$\langle \psi | \hat H |\psi \rangle$ is real on account of the Hermiticity of the operator $\hat H$. The point is that what appears in the expression \ref{psiHpsi} is not the Hilbert space inner product of the vector $ |\psi \rangle $ with the Hilbert space operator $ \hat H$ acting on $ |\psi \rangle $, but a product of the complex conjugate function with a differential operator $ H $ acting on the function, i.e. we would need to perform an integration to go from \ref{psiHpsi} to $ \langle \psi | H | \psi \rangle $, an integration which would indeed lead to a real valued quantity.}. In general, then, the recipe does not work. The conclusion seems to be that, within the pilot-wave theory, there is no satisfactory way to define a notion of energy that is generically conserved.

So far we have only considered non-relativistic pilot-wave theory. There are, however, pilot-wave approaches to quantum field theory, where a unitarily evolving wave functional is supplemented by either point particles or by fields. These additional elements are taken to evolve according to an appropriate guiding equation. The formulation of this guiding equation requires a privileged foliation of space-time, which seems to imply a violation of Lorentz invariance. In turns out, however, that this type of theories can be constructed to be fully Lorentz invariant at the \emph{empirical} level, \cite{Durretal}.

As a concrete example, consider the case of the theory of a scalar field $\phi $ on a globally hyperbolic space-time foliated in the form of $ R\times \Sigma $ with coordinates $ (x^i,t)$ adapted to the foliation (see \cite{Ward1}). The space-time metric is then described in terms of the lapse function $N$, the shift vector $ N_i$ and the induced spatial metric components $h_{ij}$ as
\begin{equation}\label{Lapse-Shift}
dS^2= -(N^2 - N_i N^i) dt^2 + 2N^i dt dx^i + h_{ij}dx^i dx^j .
\end{equation}
The functional Schrödinger equation for the wave functional $ \Psi[ \phi ]$ is then taken to be
\begin{equation}
i\frac{\partial \Psi}{\partial t} = \int_\Sigma d^3 x (N \hat { \cal H} +N_i \hat {\cal H}^i)\Psi ,
\end{equation}
where 
\begin{equation}
 \hat {\cal H} = \frac{1}{2}\sqrt{h} (-\frac{1}{h}\frac{\delta^2}{\delta \phi^2} + h^{ij} \partial_i\phi \partial_j\phi)
\end{equation}
and 
\begin{equation}
\hat {\cal H}^i = - \frac{i}{2}\sqrt{h}(\partial_i \phi \frac{\delta}{\delta \phi} + \frac{\delta}{\delta \phi} \partial_i \phi) .
\end{equation}
Then, in analogy with the particle positions of the non-relativistic case, one introduces an actual field configuration $\Phi (x, t) $ and postulates it to evolve according to
\begin{equation}
n^\mu \partial_\mu \Phi = \left. \frac{1}{\sqrt{h}} \frac{\partial S}{\partial \phi} \right\rvert_{\phi =\Phi}
\end{equation}
with $n^\mu$ the normal to the preferred foliation and $S$ the phase of the wave functional $\Psi$.
We note that, as warned, the dynamics of the field $\Phi (x, t) $ depends on the foliation.
 
The status of energy conservation does not change in the transition from non-relativistic to quantum field theoretic pilot-wave theories. As in the non-relativistic case, in the quantum field theory setting there is no satisfactory way to define a notion of energy that is generically conserved. In analogy with Eq. (\ref{EQ}), a natural energy-momentum tensor associated with the configuration variable $\Phi (x, t) $ can be defined in this case by
 \begin{equation}
T_{ab} [g, \Phi]= \partial_{a}\Phi \partial_{b}\Phi - 1/2 g_{ab} \partial^{c}\Phi \partial_{c}\Phi - \left. \frac{2}{\sqrt{-g}} \frac{\delta Q}{\delta g^{ab}} \right\rvert_{\phi =\Phi}.
\end{equation}
The problem, of course, is that such a tensor satisfies
 \begin{equation}\label{QF-Viol-pilot-wave theory}
 \nabla^{a}T_{ab} = -\frac{1}{\sqrt{-g}} \left. \left[ \frac{\delta Q}{\delta \phi } \partial_{b}\phi + 2 \nabla^{a} \frac{\delta Q}{\delta g^{ab}} \right] \right\rvert_{\phi =\Phi},
\end{equation}
so, in general, it does not have a vanishing divergence.

Next we would like to examine the semiclassical treatment of gravitation within pilot-wave theory. To do so one could, as we have done so far, source the Einstein equations with the expectation value of the energy-momentum tensor operator, $ \langle \Psi | \hat T_{ab}| \Psi \rangle $, a quantity with vanishing divergence. The problem is that such a choice overlooks the additional variables which, as we have said, are regarded by the theory as the central elements of its ontology. A much more reasonable choice in this setting would be to couple classical gravity directly to the additional variables. 

For instance, in the case of the scalar field discussed above, its energy-momentum tensor $T_{ab}$ can be directly coupled to the classical Einstein tensor
\begin{equation}
G_{a b}[g]= 8 \pi G T_{ab}[g,\Phi] .
\end{equation}

How does this approach fare regarding the issues we encountered with semiclassical gravity? We just saw that the pilot-wave energy-momentum tensor $T_{ab}$ is generically not divergenceless. Therefore, this approach does not seem to address the inconsistency problem of the semiclassical equation. The issue with the dispersion of the energy-momentum tensor, on the other hand, appears to be immediately solved because what sources the semiclassical equations is an actual field configuration, and not the expectation value of a quantum state. Let us explore in more detail each of this points.

Regarding the fact that the energy-momentum tensor of the pilot-wave field is not divergenceless, it is true that in general it is not conserved, yielding the semiclassical equations inconsistent. In \cite{Ward1}, however, a possible path to overcome this issue is proposed. The idea is that, by eliminating the gauge invariance of the theory, an extra quantum potential-dependent contribution to the energy-momentum tensor can be constructed, in such a way that the new total energy-momentum tensor is conserved. While this has not been achieved yet for the general semiclassical gravity case, it has been done for mini-superspace (i.e., symmetry-reduced) toy models and in the analogous case of pilot-wave semiclassical scalar electrodynamics (see \cite{Ward1} for details).

It is worth mentioning that the pilot-wave approach offers a natural path to deal with the full quantum gravity problem, in which both matter fields and the metric are treated quantum mechanically (see \cite{Ward2}). To do so, one again considers a globally hyperbolic space-time with a preferred foliation, characterized by a lapse function $N$ and a shift vector $ N_i$, together with a Hamiltonian characterization of the metric variables\footnote{That is, one uses the induced spatial metric on $ \Sigma_t$ , $ h_{ij}$, and its conjugate variable $ \pi^{ij} $, which is related to the extrinsic curvature $K^{ij}$.}. The wave functional $ \Psi [ \phi, h_{ij}]$ is taken to satisfy the Wheeler–DeWitt equation, which of course implies that the wave functional does not evolve in time. The interesting point is that this does not mean that the same will be the case for the actual spatial metric $H_{ij}$; such an object will undergo a nontrivial time evolution directed by the corresponding guidance equation. As a result, this approach successfully deals with the so-called \emph{problem of time} in quantum gravity.

This pilot-wave full quantum gravity scheme provides for a well-defined treatment, which has in fact been applied to a few mini-superspace examples. The price to pay, however, is the dependence of the results on the choice of $N$ \footnote{In the classical setting, the space-times resulting from different choices of lapse and shift are diffeomorphic in their regimes of validity. Therfore, they represent the same physical solution (see for instance \cite{Wald:84}).In the present context, the results are still independent of the choice of shift vector $ N_i$, but not so on the choice of lapse function $N$.}. Moreover, we must note that in this scheme the resulting space-time metric does not satisfy Einstein's equations, not even in the vacuum case\footnote{The resulting equation can (as for any well-defied evolution equation for the metric) be written in the form $G_{\mu\nu} = T{\mu\nu}$ (with the RHS being divergence free as a result of the Bianchi identity). However, in the present case, the RHS cannot be interpreted as the energy-momentum of matter, as it contains terms that do not vanish in the absence of matter.}. It would be rather interesting then to explore the degree to which solutions in this framework differ from solutions in classical general relativity. We will not discuss this path further here as it  takes us away from the semiclassical setting on which we want to focus.

Above we mentioned that the pilot-wave approach appears to instantly solve the problem with semiclassical gravity associated with the fact that states with a large dispersion in energy yield the theory empirically inadequate. This is because, in the pilot-wave approach, what sources the semiclassical equations is an actual field configuration, and not the expectation value of a quantum state. That is, even if the quantum state guiding the field configuration contains a large dispersion in energy, the actual field variable will have a well-defined configuration. This, however, does not necessarily mean that the theory will yield prediction in accordance with observations. To explore this issue in more detail, let us analyze the gedankenexperiment in the pilot-wave context. In such an experiment, as the two paths available to the photons separate in space, the wave function becomes a sum of components with support in two disjoint regions of configuration space. The pilot-wave photons will then be in either one of them and in each case they will have a well-defined energy assignment. However, the point of the set up is precisely that the wave components will eventually rejoin in configuration space, leading to a superposition of the components corresponding to the two paths. At such a point, the behavior of the pilot-wave photons will be governed by the superposition and will cause the photons to behave in a non-standard fashion. The crucial question is if the space-time generated by such a behavior via the pilot-wave semiclassical gravity equations is compatible with experiments. 

The conclusion of this analysis seems to be that the pilot-wave theory does not straightforwardly lend itself to a peaceful coexistence with semiclassical general relativity. At this point it is not clear how serious the departures from that would be in some concrete situations and to what degree some kind of approximated treatment would be viable. To our knowledge, with the exception of the initial analysis made in \cite{Ward1,Ward3}, the issue remains generally unexplored.

\subsubsection{Everettian frameworks}

Everettian frameworks attempt to explain why observers obtain measurement results in accord with the standard quantum statistics \emph{in spite of} postulating that the wave function is complete and always evolves according to the Schrödinger equation (i.e., without invoking hidden variables or actual, physical collapses of the quantum state). To achieve this, they propose that the different terms of a macroscopic superposition must be interpreted as describing a \emph{multiplicity} of almost non-interacting macroscopic ``worlds’’. In particular, they maintain that all possible results of a quantum experiment obtain, each of them in a different branch or world. In contemporary Everettian frameworks, these different worlds are local, macroscopic, emergent phenomena that arise from decoherence caused by the interaction between systems and their environment \cite{WallaceBook}.

Everettian frameworks seem elegant and compelling. However, despite of intense work on the subject, they might still face open issues, particularly with respect to ontological matters and the role that probability plays in the theory. We will not consider in this paper whether these issues have been dealt with satisfactorily. We will restrict our attention to exploring the status of conservation laws within this type of frameworks.

Since in the Everettian approach there are no quantum collapses or hidden variables, it might seem that there are no problems with energy conservation. All there is, is a unitarily evolving wave function, out of which a conserved energy appears to be definable. Things, however, are not that simple. One could, for instance, define energy via the EE rule. However, if so, there are two options: either the state of the universe is an eigenstate of the Hamiltonian, in which case energy is conserved, but nothing ever changes, or the state of the universe is not an eigenstate of the Hamiltonian, in which case the value of the energy of the universe is never well-defined.

Alternatively, one could define total energy as the expectation value of the Hamiltonian, which is always well-defined and (if the Hamiltonian is time-independent) conserved. The question, though, is how useful is this definition. The problem is that such a definition is associated with a \emph{global} perspective, pertaining to the whole wave function, which does not describe our experience at all. Such an experience must be extracted from the multiplicity of branches that coexist in the world. For instance, we would want to talk about the energy of a specific particle we study in an experiment, or to talk about, say, the physical geometry of the world around us, i.e., in the branch of the world that we can empirically explore and which determines our experiences.

What can be said, then, about energy conservation within each branch? To explore the issue, consider again an energy measurement on a system $S$ in the state 
\begin{equation}
|\psi\rangle_S = \frac{1}{\sqrt{2}} \left( |E_1\rangle_S + |E_2\rangle_S \right) ,
\end{equation}
with $|E_1\rangle_S$ and $|E_2\rangle_S$ energy eigenstates with $E_1 \neq E_2$. According to the Everettian approach, after the measurement there will be two branches, one in which the result is $E_1$ and one in which it is $E_2$. Clearly, in both branches, the expectation value of the Hamiltonian is \emph{not} conserved. That is, even if the wave function of the world evolves unitarily, each branch experiences an \emph{effective} evolution which, at least at times, deviates from unitarity---and thus implies energy non-conservation.

As in the standard scenario, one could argue that if one takes into account not only $S$, but also the measuring apparatus, the environment, etc., then energy would be conserved within each branch. This, however, would require the branching structure to always align with the eigenvectors of the Hamiltonian, which would imply no time evolution within the branches---a conclusion clearly at odds with experience. More generally, it seems that all the problems we encountered within the standard framework regarding energy conservation carry over verbatim to the case of the effective evolution within each branch. It is important to notice, though, that as with the other approaches we have considered, this lack of energy conservation does not lead to internal inconsistencies.

Things become much more interesting as soon as one adds gravity to the picture, particularly if one enquires about the relation between space-time and matter in an Everettian scenario. One option would be to assume that all branches share the same space-time structure, sourced by the expectation value of the energy-momentum distribution of the universal quantum state. Since such a state evolves unitarity, the energy-momentum expectation value is always divergenceless and the semiclassical Einstein equations are fully consistent. The problem, of course, it that this option is shown to be empirically inadequate by the experiment in \cite{PagGei:81}. It seems necessary, then, to assume that each branch has its own space-time\footnote{In \cite{Kent} another option is considered which, at first sight, seems unattractive: to assume that all branches share the same space-time, sourced by the matter distributions of one of the branches. \cite{Kent} argues that, by adopting a gravity-based psychophysical parallelism, it is possible to solve some of the obvious problems of this alternative. It is shown, however, that serious complications remain in this path.}. Now, since within each branch the matter distribution follows an effective dynamics which is not always unitary, the expectation value of the energy-momentum tensor in each branch is not always \emph{effectively} divergenceless and the semiclassical Einstein equations fail to be consistent. That is, the unitary evolution of the universal wave function within Everettian frameworks does not solve the self-consistency issue within semiclassical gravity. Is this the end of Everettian, semiclassical gravity? Well, at this stage one could employ the same defense we sketched within the objective collapse discussion: one could see the semiclassical equations as an effective description, with a limited (albeit large) range of applicability. That is, one could assume the semiclassical equations to approximately hold most of the time, but to completely break down when the effective evolution significantly deviates from unitarity at the branching ``events'' and the expectation value of the energy-momentum tensor of each branch is not divergenceless.

Above we pointed out that, even when adopting this non-fundamental reading of the semiclassical equations, they run the risk of becoming empirically inadequate if the state, within each branch in this Everettian case, contains a large dispersion in energy and momentum. We claimed that objective collapse models and the pilot-wave approach contain means to potentially alleviate this issue. In the case of collapse models, in particular, we argued that by choosing as collapse operators the components of the energy-momentum tensor---an attractive choice due to several independent reasons---one guarantees the dispersion in energy and momentum to generically remain small. What can we say about this issue within Everettian frameworks? At this point, the unresolved issues of the approach threaten to block a rigorous discussion of this matter. Without a detailed description of the mechanism that fixes the branching structure, it is hard to ascertain the degree of dispersion in energy and momentum within each branch. Of course, it could be that the branching structure brought about by, say, decoherence, generically holds such dispersions small. However, at least at first sight, there does not seem to be a mechanism that ensures that to be the case.

To see this, consider again the gedankenexperiment of section \ref{qrel}. Within an Everettian framework, one would expect that if the number of photons is sufficiently large---large enough to produce a dispersion of energy that would cause empirical problems for the semiclassical equations---then decoherence would kick in, leading to a branching to two worlds, one in which the photons have energy $E$ and one in which they have energy $ \chi \chi' E $. It seems, then, that through this mechanism problematic dispersions of energy and momentum would be avoided. The problem, though, is that the gedankenexperiment, by construction, leads to a superposition of two macroscopic scenarios, each containing a well-defined value of \emph{energy}. However, that does not always have to be the case: nothing within the decoherence mechanism protects the different branches of containing arbitrarily large dispersions in energy or momentum. Moreover, and in contrast with collapse models, which have ``moving parts'' (such as the collapse operator or the value of the collapse rate) that could be played with to address the issue at hand, Everettian frameworks are pretty much fixed, and therefore seemingly unable to react and resolve the problem. 

In sum, to the extent at which conservation laws could be said to hold within the Everettian approach, they do so in connection with the complete wave function, involving simultaneously all branches and all possible experimental outcomes. This might be taken as offering some metaphysical relief, but does little to remove the stark conclusion that the world that we have access to is one where neither conservation laws nor any exact version of Einstein's equations hold.

\section{Conclusions}

We explored the status of conservation laws in classical and quantum physics. In the classical context we revisited the transition from an ``energy as a substance'' viewpoint to the symmetry-conservation link discovered by Noether. Under the latter, we found conservation laws to be useful when available (or at least approximately available), but often not essential. Still, we noted that, in the context of general relativity, divergenceless of energy-momentum is necessary to ensure the self-consistency of the theory.

In the quantum setting, we found that the ambiguous nature of the standard formalism precludes a rigorous analysis of the conservation issue. We also explored the matter in the context of objective collapse models, pilot-wave theory and Everettian frameworks. In all cases we found no satisfactory way to define a (useful) notion of energy which is generically conserved. 

Next we explored the consequences of the energy conservation analysis for the semiclassical gravity program. We first pointed out two potential problems for such an approach: a threat to self-consistency due to energy-momentum non-divergenceless and a threat to empirical adequacy due to a possibly large energy-momentum dispersion. Regarding the first issue, we explained how a non-fundamental reading of the semiclassical framework partially alleviates the problem.

We then examined how the different quantum approaches fared in the semiclassical setting. Objective collapse models had not much to say regarding the non-divergenceless of the energy-momentum tensor but were found promising in preventing the occurrence of arbitrarily large dispersions in energy or momentum. Pilot-wave theory, on the other hand, could potentially solve the non-divergenceless issue by the inclusion of a suitable quantum current (as it has been done in \cite{Ward1} for semiclassical electrodynamics). As for the dispersion issue, the presence of additional configuration variables allows the pilot-wave approach to divert it. The price to pay, however, is the need for a preferential foliation and the fact that, generically, spacetime fails to satisfy Einstein's equations. Everettian frameworks do not seem to offer concrete tools to deal with any of the two potential problems with the semiclassical approach. It seems that we must accept the fact that neither conservation laws nor Einstein's equations hold in the world we actually experience.
\begin{flushright}
•
\end{flushright}
\section*{Acknowledgments}
We are happy to acknowledge very interesting discussions with, and multiple helpful comments from Ward Struyve. DS work was supported by a sabbatical fellowship from PASPA-DGAPA-UNAM, by Fulbright -Garcia-Robles -COMEXUS- fellowship, and by the grant FQXI-MGA-1920 from the Foundational Questions Institute and the Fetzer Franklin Fund, a donor advised by the Silicon Valley Community Foundation. EO acknowledges partial financial support from PAPIIT-UNAM grant IN102219.
\bibliographystyle{plain}
\bibliography{biblio.bib}
\end{document}